\documentclass[10pt, conference, letterpaper]{IEEEtran}

\usepackage{multicol}
\usepackage{multirow}
\usepackage{graphicx}

\usepackage{cite}
\usepackage{url}

\usepackage{amssymb}
\usepackage{amsthm, amsmath}
\usepackage{amsbsy}
\usepackage{color}

\usepackage{subfigure}

\usepackage{cite,url}

\usepackage{algorithmic}
\usepackage[ruled]{algorithm2e}
\SetKwInput{kwInitStep}{Initialization Step}
\SetKwInput{kwGibbsStep}{Gibbs Sampling Step}
\SetKwInput{kwTrainStep}{Training}
\SetKwInput{KwPredictStep}{Prediction}

\def\squareforqed{\IEEEQED}
\def\qed{\ifmmode\squareforqed\else{\unskip\nobreak\hfil
\penalty50\hskip1em\null\nobreak\hfil\squareforqed
\parfillskip=0pt\finalhyphendemerits=0\endgraf}\fi}

\newcommand{\ud}{\mathrm{d}}

\IEEEoverridecommandlockouts
\begin{document}

\title{
	House Price Modeling over Heterogeneous Regions with Hierarchical Spatial Functional Analysis
}

\author{\IEEEauthorblockN{Bang Liu\textsuperscript{*}\thanks{\textsuperscript{*}These authors contributed equally to this work.}\thanks{ This work was partially supported by the NSERC-CRD and NSERC-RGPIN grants.}}
\IEEEauthorblockA{
University of Alberta\\
bang3@ualberta.ca}
\and
\IEEEauthorblockN{Borislav Mavrin\textsuperscript{*}}
\IEEEauthorblockA{
University of Alberta\\
mavrin@ualberta.ca}
\and
\IEEEauthorblockN{Di Niu}
\IEEEauthorblockA{
University of Alberta\\
dniu@ualberta.ca}
\and
\IEEEauthorblockN{Linglong Kong}
\IEEEauthorblockA{
University of Alberta\\
lkong@ualberta.ca}}

\maketitle
\thispagestyle{empty}
\pagestyle{empty}

\begin{abstract}
	Online real-estate information systems such as Zillow and Trulia have gained increasing popularity in recent years. One important feature offered by these systems is the online home price estimate through automated data-intensive computation based on housing information and comparative market value analysis. State-of-the-art approaches model house prices as a combination of a latent land desirability surface and a regression from house features. However, by using uniformly damping kernels, they are unable to handle irregularly shaped regions or capture land value discontinuities within the same region due to the existence of implicit sub-communities, which are common in real-world scenarios. In this paper, we explore the novel application of recent advances in spatial functional analysis to house price modeling and propose the Hierarchical Spatial Functional Model (HSFM), which decomposes house values into land desirability at both the global scale and hidden local scales as well as the feature regression component. 
We propose statistical learning algorithms based on finite-element spatial functional analysis and spatial constrained clustering to train our model.
Extensive evaluations based on housing data in a major Canadian city show that our proposed approach can reduce the mean relative house price estimation error down to $6.60\%$.
\end{abstract}
\section{Introduction}
\label{sec:intro}

Online real-estate information systems, such as Trulia, Zillow, Yahoo! Homes and Realtor.com, have gained enormous popularity in a trend termed as online-to-offline commerce (O2O). One important feature offered by these online systems, e.g., Zestimate \cite{zestimate}, Trulia Estimates \cite{trulia}, is the estimate of a home's worth through machine learning based on diverse real-estate information, including comparative sales of similar homes and home attributes like the numbers of bedrooms and bathrooms, square footage, lot sizes, build year, characteristics of the basement, presence of fireplaces or air-conditioning, etc. For example, about 100 million homes across the US currently have Zestimate values \cite{zestimate}. Although automatically computed home price estimates may not substitute official appraisals conducted by banks, governments or real-estate agents, they serve as a starting point in determining home values, provide vast amounts of valuable data to homebuyers and investors, and can greatly aid offline real-estate transaction decisions. 

House price estimation is an interdisciplinary research topic that has been widely studied by researchers from diverse backgrounds, including economists, statisticians and computer scientists. Early methods are based on parametric models, including Hedonic models \cite{Dubin98,Bourassa99} based on linear regression of house attributes and CS index based on repeat sales modeling \cite{Case89}. 
Later models take into account spatial correlation \cite{Dubin98, Pace98} between houses in a same community or region to estimate a house's market value based on neighboring houses with known values, e.g., those that are sold in the recent year. 

A milestone in modern data-intensive real-estate valuation literature is 
the introduction of semi-parametric or non-parametric models \cite{Clapp04, Chopra07, Caplin08}, which assume that the house prices can be decomposed into a latent underlying land desirability and a regression of house attributes. In all these contemporary semi/non-parametric models, a common assumption is that the underlying land desirability of a house is a weighted average of its neighbours' desirability, where the weights as well as the neighbourhood of the house are determined by certain kernel functions used to assess the closeness of two houses in the feature space.
However, such kernel-based semi/non-parametric methods suffer from several intrinsic weaknesses. First, its hard to choose the specific form of kernel functions with a large number of hyper-parameters to tune. More importantly, by using uniformly damping weights depending on the distance, kernel functions cannot handle irregularly shaped non-convex regions or regions with interior uninhabited areas, which are common in reality. For example, when calculating the latent desirability of a house, a ``neighbour'' on the other side of a river may have the same influence as a ``neighbour'' on the same side of the river, as long as the two ``neighbours'' have the same distance to the house in the feature space. Neither can they handle land desirability discontinuities within a same region containing implicit sub-communities, e.g., estate homes vs. regular homes, homes near a coast or on a hill with views vs. other homes, etc. 

In this paper, we explore the novel application of latest advances in spatial functional learning and finite element analysis to real-world house price modelling problems. We propose the \emph{hierarchical spatial functional model} (HSFM), which can handle irregularly shaped geographic regions as well as model land desirability discontinuities within heterogenous regions in reality. We have made multiple contributions:

\emph{First}, similar to kernel-based methods, we assume the house value is composed of a regression from house features and a latent land desirability surface that depends on location. However, in our spatial functional approach, a smooth latent surface is learnt through a finite-element spatial spline regression approach recently developed in statistics \cite{Sanga13, wood2003thin, Ramsay02}, thus overcoming the limitation of uniform damping in kernel-based methods and being able to handle any irregularly shaped non-convex regions.

\emph{Second}, the original spatial spline regression still yields large errors in modeling real-world house price data, as it tends to smoothen a spatial field throughout the region of interest, which may actually be heterogeneous in terms of both house prices and features in reality, characterized by implicit patches and hidden land value discontinuities. In our hierarchical spatial functional model, to further handle hidden land value discontinuities, we assume that a home's value is the combination of the \emph{global} land desirability in the region of interest, and the \emph{local} land desirability in the hidden sub-region or sub-community it belongs to. We develop an effective method to learn this hierarchical model through the use of spatial constrained partitioning based on locations and the latent land desirability, to automatically discover hidden sub-communities. 

We evaluate the proposed approach on a dataset of over 6000 houses in Edmonton, Alberta, Canada in a typical region along a river, featuring heterogeneous hidden sub-communities and strong non-convexity. Results suggest that our HSFM approach  
reduces the root mean square error (RMSE) by 28.22\% and reduces the mean relative absolute error (MRAE) by 21.61\% over original spatial spline regression,
leading to an MRAE of only 6.60\% in house price estimation.

\section{Relationship to Prior Work}
\label{sec:related}

House market value estimation in a region based on partially known house prices is an interdisciplinary research topic which lies at the intersection of urban economics, statistics and computer science. 

\textbf{Parametric Models.} 
Earlier methods are mainly based on linear regression. The so-called hedonic model says that the price of a house is a linear function of its attributes \cite{Dubin98}, \cite{Bourassa99}, e.g., lot size, square footage, number of bedrooms, distance to landmarks, etc. However, the method is limited by the linearity assumption, and it is impossible to gather all relevant attributes. One important development in real-estate value estimation is to consider the spatial correlation,
since many implicit community variables, e.g., employment rate, income, school, land desirability, etc., are hard to measure but will lead to spatial correlation of house prices. 
\cite{Dubin98}, \cite{Pace98}
incorporate the spatial autocorrelation into the residuals of the linear regression hedonic model.
However, such parametric models all assume a fixed number of parameters and cannot grow the number of parameters with the amount of training data. This becomes a serious limitation in the presence of an ever-increasing amount of data.




\textbf{Non-parametric Models and Kernel-Based Methods}. 
Semi/non-parametric models have appeared in modern computational-intensive house valuation techniques, and almost all existing methods of this kind fall in the class of nearest neighbour models and adopt a kernel-based interpolation \cite{Pace93}, \cite{Anglin96} or the so-called kernel-based locally weighted regression \cite{Clapp04}, \cite{Caplin08}. However, the major limitation of kernel-based methods mentioned above is that the kernel function used to dictate the weights of correlation between neighboring houses $i, j$ is uniformly damping with their distance $\|X_i-X_j\|$ in the feature space. Due to this reason, kernel functions is not suitable handle non-convex regions, since they tend to link houses across uninhabited areas (e.g., a river or a hill) where the linkage is weak. Neither can kernel functions model complex phenomena such as sudden land value jumps within the same region due to the existence of sub-communities. 


\textbf{Spatial Functional Analysis}.
Spatial functional analysis \cite{Sanga13} has been applied to census information prediction and temperature prediction. It is able to recover a smooth spatial field over non-convex regions with irregular boundary.
To the best of our knowledge, we are the first to apply latest spatial functional analysis techniques to real-estate price modeling, assuming the house price consists of a regression part based on house features and a spatial field that models the land desirability at a certain location. 
In particular, we leverage the idea of spatial spline regression to handle any irregularly shaped geographic regions and thus overcome the limitations of the existing kernel-based methods in house valuation literature. 
Furthermore, we have developed a novel hierarchical spatial functional model to interpret the inherent house price heterogeneity that exists in most real-world regions of interest, and propose effective training algorithms to automatically capture land desirability jumps between the hidden sub-communities that are common in reality.




\section{Models}
\label{sec:model}

The problem of house value estimation through comparative analysis is to estimate the market values of all houses in a certain region, based on information like locations, attributes, and the values of a subset of all houses (e.g., those recently sold).
Suppose the region of interest is modelled by an irregularly bounded domain $\Omega \subset\mathbb R^2$ that excludes the uninhabited areas such as parks, rivers, industrial areas, hills and so on.
$\Omega$ is given as an input and can be generated from the collection of houses being considered according to a process to be outlined in Sec.~\ref{sec:algorithm}. 
Let $\{\mathbf p_i = (x_i,y_i)\}\in \Omega$ denote the geographic location (e.g., GPS coordinate) of each property $i$ in the bounded domain. Let $z_i$ be the value (or log value) of property $i$ and $\mathbf w_i = (w_{i1},\ldots,w_{iq})^\mathsf{T}$ be a vector of $q$ covariates associated with property $i$ that represents $q$ attribute values (e.g., lot size, net area, number of bedrooms/bathrooms/garages, age, etc.) 

\begin{figure}[t]
		\centering
        \includegraphics[width=2.8in]{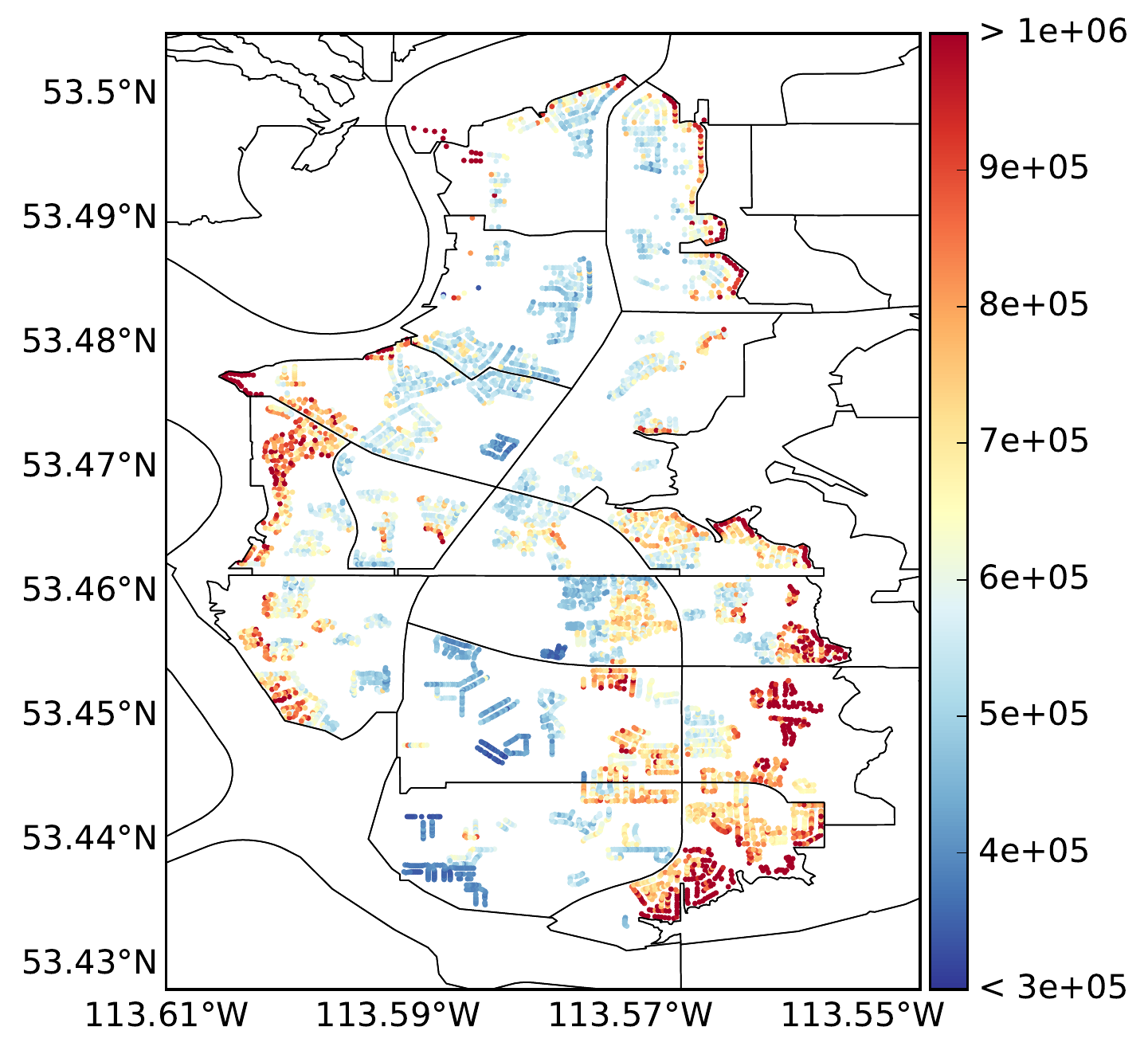}
        \caption{House prices (CAD) of Riverbend area in Edmonton with North Saskatchewan River on the left (with an uninhabited park at $53.48\,^{\circ}\mathrm{N}$) and Whitemud Creek ravine on the right. Mean: CAD 655,894, median: CAD 598,000, min: 65,000, max: CAD 2,855,500. The dark lines denote the boundaries of municipal communities. }
        \label{fig:region}
\vspace{-5mm}
\end{figure}

\subsection{The Spatial Functional Model}

Modern house value estimation literature \cite{Clapp04, Chopra07, Caplin08} all models the value of a house as a combination of its underlying land desirability and attribute value, leading to a semi-parametric model:
\begin{equation}\label{eq:spatial_model}
	z_i = {\mathbf w}_i^{\mathsf T}\boldsymbol{\beta} + f(\mathbf p_i) + \epsilon_i,
\end{equation}
where $f$ is a real-valued function or a spatial field that models the underlying land desirability over different positions $\mathbf p_i\in\Omega$ and $\boldsymbol{\beta}\in\mathbb R^q$ contains the regression coefficients for the property's attributes. The scalars $\epsilon_i$ are independent residuals.
When $z_i$ is the original house value, \eqref{eq:spatial_model} is an additive model. On the other hand, when $z_i$ is the log value of house $i$, \eqref{eq:spatial_model} is essentially a multiplicative model for the original house value and can accommodate multiplicative factors, e.g., lot size, square footage, in an additive form \cite{Case89, Clapp04, Chopra07, Caplin08}.

Furthermore, $f$ is assumed to be twice-differentiable over $\Omega$. The model above enables a key assumption that many spatial data studies hinge upon, that is, the property lots close to each other are likely to have a similar land desirability, while price variations among neighbours are still captured by different attributes ${\mathbf w}_i$.

Both the land desirability $f$ and the covariate coefficients $\boldsymbol{\beta}$ can be learned based on $n$ training data points in $\Omega$, which contain the following information: \emph{1)} the values (or log values) of these $n$ points: $\mathbf z := (z_1, ... , z_n)^{\mathsf T}$, \emph{2)} their positions $\{\mathbf p_i: i = 1,\ldots,n\}$, and \emph{3)} an $n \times q$ covariate matrix $\mathbf W$, whose $i$th row is given by $\mathbf w_i^{\mathsf T}$.
Once $f$ and $\boldsymbol{\beta}$ are learned from the training data, the value $z_j$ of any property $j$ can be estimated by plugging its attribute information ${\mathbf w}_j$ and position $\mathbf p_j$ into \eqref{eq:spatial_model}.
The model \eqref{eq:spatial_model} can be trained using either kernel-based methods \cite{Clapp04, Chopra07, Caplin08} or finite element analysis \cite{Sanga13, wood2003thin, Ramsay02}.

However, the weakness of model \eqref{eq:spatial_model} is that it is not capable  of capturing the inherent heterogeneity that exists in real-world regions. Fig. \ref{fig:region} shows that the house prices of Riverbend area in Edmonton, AB Canada. The house prices in the middle-left region and the bottom-right corner are much higher than other regions. It turns out that the North Saskatchewan River is on the left boundary and the middle-left area is also next to the Terwillegar Park. Therefore, the houses in that region have more privacy and better views. The bottom-right corner has similar profile as it is near the Whitemud Creek. Moreover, the houses in this area were built more recently and thus have more expensive interior and exterior features than others. 

Such hidden heterogeneity related to house features and/or land values leads to the discontinuity/jump of land desirability surface and varied attribute coefficients. These sudden jumps in desirability and attribute coefficients across local subregions can not be captured by model \eqref{eq:spatial_model}, where a smooth $f$ and a constant $\boldsymbol{\beta}$ are assumed over $\Omega$. Furthermore, such local subregions are usually hidden and are not explicitly outlined by the municipal division, as shown in Fig.~\ref{fig:region}.

\subsection{A Hierarchical Spatial Functional Model}
\label{sec:submodel}
We now propose the \emph{hierarchical spatial functional model} (HSFM) to learn the values (or log values) $z_i$ by the sum of a global value function $g$ and a local value function $l$, both depending on attributes $\mathbf w_i$ and locations $\mathbf p_i$. To learn both value functions $g$ and $l$, we adopt the semi-parametric model in \eqref{eq:spatial_model}. In summary, our hierarchical spatial functional model is of the form
\begin{align}\label{eq:hier_model}
	z_i &= g({\mathbf w_i},\mathbf p_i) + l({\mathbf w_i},\mathbf p_i) +\epsilon_i,\\ \nonumber
	g({\mathbf w}_i,\mathbf p_i) &= {\mathbf w}_i^{\mathsf T}\boldsymbol{\beta}^g + f^g(\mathbf p_i) + \epsilon_i^g,\\ 
	  l({\mathbf w}_i,\mathbf p_i)& = \sum\nolimits_{i\in \Omega_j}\left( {\mathbf w}_i^{\mathsf T}\boldsymbol{\beta}^{l_j} + f^{l_j}(\mathbf p_i)\right) +\epsilon_i^l, \nonumber
\end{align}
where $\Omega_j \cap \Omega_{j'} = \emptyset$, for $j\neq j'$,  $\cup_j \Omega_j = \Omega$, $j=1,\ldots,J$, which means that $\{\Omega_j,~j=1,\ldots,J\}$  is a partition of $\Omega$ and the scalars $\epsilon_i$, $\epsilon_i^g$, and $\epsilon_i^l$ are independent residuals. The global value function $g$ is homogeneous in the entire domain $\Omega$. The local value function $l$ is homogenous in each local region $\Omega_j$ yet heterogeneous between different local regions $\Omega_j$. In other words, $\boldsymbol{\beta}^g$ captures the global contributions of attributes $\mathbf w_i$, while $\boldsymbol{\beta}^{l_j}$ capture the local attribute contributions and  are different  for each local region $\Omega_j$. Similarly,  the spatial field $f^g$ represents the global underlying land desirability, while the spatial fields $f^{l_j}$ stand for the local underlying land desirabilities, which are different for each local region. 

Our hierarchical spatial functional model has two levels, a global level and a local level. It is more general than \eqref{eq:spatial_model} where only a global level is learned and becomes the same as \eqref{eq:spatial_model} without the local value function $l$. At the global level, our model learns the global coefficients $\boldsymbol{\beta}^g$ for attributes ${\mathbf w_i}$ and the global spatial field $f^g$. At the local level, the local attribute coefficients  $\boldsymbol{\beta}^l$ and local spatial fields $f^{l_j}$ are learned in each local homogeneous subregion. Therefore, our model is more powerful than model \eqref{eq:spatial_model} and is capable of not only characterizing the global trends but also distinguishing the local variabilities to handle land desirability discontinuities. There are various methods to identify the homogeneous local regions $\Omega_j$.
In this work, we use a spatial constrained clustering method based on the global land desirability $f^g$ to partition the whole region $\Omega$ into subregions, as shown in Fig. \ref{fig:cluster3}. 

The model \eqref{eq:hier_model} can be learned in a sequential way like other hierarchical models.  In particular, we first estimate the global value function $g({\mathbf w}_i,\mathbf p_i)$ as $\hat g({\mathbf w}_i,\mathbf p_i)$, based on the training data, which can be done using the same methods for learning model \eqref{eq:spatial_model}. Then, using the residuals $z_i - \hat g({\mathbf w}_i,\mathbf p_i)$, we obtain the estimate of the local value function $\hat l({\mathbf w}_i,\mathbf p_i)$ for each local region based on certain partitions of the whole region $\Omega$. Note that the partitions may depend on our estimated global spatial fields $\hat f^g(\mathbf p_i)$ in the first step. We describe our learning algorithms in details in Sec.~\ref{sec:algorithm}. 



\section{Learning Algorithms}
\label{sec:algorithm}

In this section, we first describe a finite-element analysis approach called Spatial Spline Regression (SSR) that can approximately solve model \eqref{eq:spatial_model} for both $f$ and $\boldsymbol{\beta}$ over any irregularly shaped domain $\Omega$. We then present the training algorithm of our hierarchical spatial functional model \eqref{eq:hier_model} to learn $f^g$ and $\boldsymbol{\beta}^g$ at the global scale as well as each $f^{l_j}$ and $\boldsymbol{\beta}^{l_j}$ in local regions.

\subsection{Spatial Spline Regression}

Spatial spline regression \cite{Sanga13} is a recently proposed finite-element analysis approach to jointly solve for $f$ and $\boldsymbol{\beta}$ from model \eqref{eq:spatial_model} over any irregularly shaped domain $\Omega$. To obtain a smooth estimate of the land desirability spatial field $f$, based on functional analysis, the following penalized sum of square errors is minimized \cite{Sanga13}, \cite{Ramsay02}:
\begin{equation}\label{eq:min_err}
	\text{minimize}\ \mathcal L( \boldsymbol{\beta}, f) = \sum_{i=1}^{n}\big(z_i- {\mathbf w}_i^{\mathsf T} \boldsymbol{\beta} - f(\mathbf p_i) \big)^2 + \lambda\int_\Omega(\nabla^2 f)^2\ud\mathbf p,
\end{equation}
where $\nabla^2 f= \frac{\partial^2 f}{\partial x^2}+\frac{\partial^2 f}{\partial y^2}$ denotes the Laplacian of $f$ over $\Omega$ to smoothen out the roughness of the spatial field $f$, since neighboring houses should share similar land desirability. The tuning parameter $\lambda$ is used to control the smoothing of the land value surface and can be selected using some data-driven or {\it ad hoc} methods. 

However, the challenge to solving problem \eqref{eq:min_err} is that it involves searching for a functional $f$ over a non-convex domain $\Omega$ that may have strong concavities, complicated boundaries, and even interior holes. 
Spatial spline regression \cite{Sanga13} can solve the type of problem \eqref{eq:min_err} via finite element analysis: the domain $\Omega$ is divided into small disjoint triangles, and a polynomial function is defined on each of these triangles, such that the summation of these polynomial functions defined on different pieces closely approximates the desired $f$. In our data analysis, we use the package \emph{Triangle}  (version 1.6) to generate Delaunay triangulation mesh for finite element analysis, see \url{https://www.cs.cmu.edu/~quake/triangle.html} for details. 

Specifically, let $\zeta_1,\ldots,\zeta_K$ denote the vertices of all the small triangles, which are called control points and can be adaptively selected by available data points. We define a piecewise linear or quadratic basis function $\psi_k(x,y)$ called {\it Lagrangian finite element} with $(x,y)\in\Omega$, associated with each control point $\zeta_k$ such that $\psi_k$ evaluates to $1$ at $\zeta_k$ and is equal to $0$ at all other control points. Therefore, according to the {\it Lagrangian property of the basis} we can approximate $f(x,y)$ for any $(x,y)\in\Omega$ only using the values of $f$ on the $K$ control points, i.e., $\mathbf f := (f(\zeta_1),\ldots,f(\zeta_K))^{\mathsf T}$. That is,
if we let $\psi(x,y) :=(\psi_1(x,y),\ldots,\psi_K(x,y))^{\mathsf T}$ denote the $K$ predefined basis functions, each corresponding to a control point, then we have
\begin{equation}\label{eq:triApprox}
	f(x,y) = \mbox{$\sum_{k=1}^{K}$} f(\zeta_k)\psi_k(x,y) = \mathbf f^{\mathsf T}\mathbf\psi(x,y),
\end{equation}  
Since $\psi_1(x,y),\ldots,\psi_K(x,y)$ are predefined and known \emph{a priori}, the variational estimation of $f$ in problem \eqref{eq:min_err} boils down to the estimation of only $K$ scalar values, i.e., $\mathbf f = (f(\zeta_1),\ldots,f(\zeta_K))^{\mathsf T}$.
With the piece-wise approximation given by \eqref{eq:triApprox}, solving \eqref{eq:min_err} is simply solving a set of linear equations for $\hat f(\zeta_1),\ldots,\hat f(\zeta_K)$. 
The estimator $\hat f(x,y)$ for $f$ can then be derived from \eqref{eq:triApprox}, as shown in  \cite{Sanga13}.


 


\subsection{Learning the Hierarchical Spatial Functional Model}

\begin{algorithm}
\caption{Hierarchical Spatial Functional Model}\label{alg:HSFM}
\KwIn{The $n$ training data points $S_{\text{train}}=\{(z_i, \mathbf{w}_i, \mathbf{p}_i)|i=1,\ldots, n\}$, smoothing parameter $\lambda$, and the domain $\Omega$ of interest. }
\KwOut{Global spatial field and parameters $\hat{f}^g,\ \hat{\boldsymbol{\beta}}^g$; local spatial fields and parameters $\{\hat{f}^{l_j},\ \hat{\boldsymbol{\beta}}^{l_j}|j =1,\ldots, J\}$.}

\begin{algorithmic}[1]
	\STATE Fit the global model $\hat{f}^g, \hat{\boldsymbol{\beta}}^g$ by applying SSR on $S_{\text{train}}$ over the entire region $\Omega$ and obtain
	the residuals: $r_i \leftarrow z_i - \mathbf{w}_i^{\sf T}\hat{\boldsymbol{\beta}}^g - \hat{f}^g(\mathbf{p}_i)$.
	
\STATE Partition the $n$ training data points into $J$ local regions
using the spatial constrained clustering approach (refer to Sec.~\ref{sec:algorithm1}) based on the estimated global field $\hat{f}^g$. 

\FOR{each cluster $j= 1,\ldots,J$}
  \STATE Fit the local model $\hat{f}^{l_j}, \hat{\boldsymbol{\beta}}^{l_j}$ with SSR on data points of each local cluster, i.e., $S_{\text{train}}^j = \{(r_i, \mathbf{w}_i, \mathbf{p}_i)|i =1,\ldots,n_{l_j}\}$, over each sub-region $\Omega_j$, where $n_{l_j}$ is the number of training points in cluster $j$.
\ENDFOR

\end{algorithmic}
\end{algorithm}

Although spatial spline regression is capable of handling irregularly shaped regions, its main disadvantage is that it tends to smoothen the land desirability throughout an entire region of interest  and can not capture price discontinuities between hidden sub-communities that are exist in the heterogeneous region, e.g., the left lower corner region in Fig. \ref{fig:region}. Hierarchical Spatial Functional Model (HSFM) solves this issue by performing spatial functional analysis at both the global scale and local scales for each hidden sub-region to characterize the price jumps.

We use a cascaded approach to train an HSFM in Algorithm~\ref{alg:HSFM}. First, a global spatial field $f^g$  and global attribute contributions ${\boldsymbol{\beta}}^g$ are fitted for the entire region $\Omega$ using SSR.
This removes the overall global trend and obtain the global residuals.
Second, we generate a geo-partition of the entire region $\Omega$ containing $\Omega_j$, $j=1,\ldots,J$, using a \emph{spatial constrained clustering} algorithm to be described in detail in the next subsection. The key idea and the challenge of our geo-partition algorithm are to obtain clusters that are both geographically separated and characterized by price discontinuities between clusters.
Third, for each local region $\Omega_j$, we learn a local spatial field $f^{l_j}$  and local attribute contributions ${\boldsymbol{\beta}}^{l_j}$ based on the global residuals of the points in $\Omega_j$ to capture the local information. Since the local spatial fields are fitted for each $\Omega_j$ disjointly, the hierarchical method allows for the non-smoothness or discontinuities between the local patches $\Omega_j$, $j=1,\ldots,J$. As a result, this sequential learning approach significantly increase the prediction power by exploring the intrinsic hierarchical structure of house prices. 


\subsection{Geo-Partitioning Algorithm}
\label{sec:algorithm1}

\begin{figure}[t]
 \centering
 \includegraphics[width=2.5in]{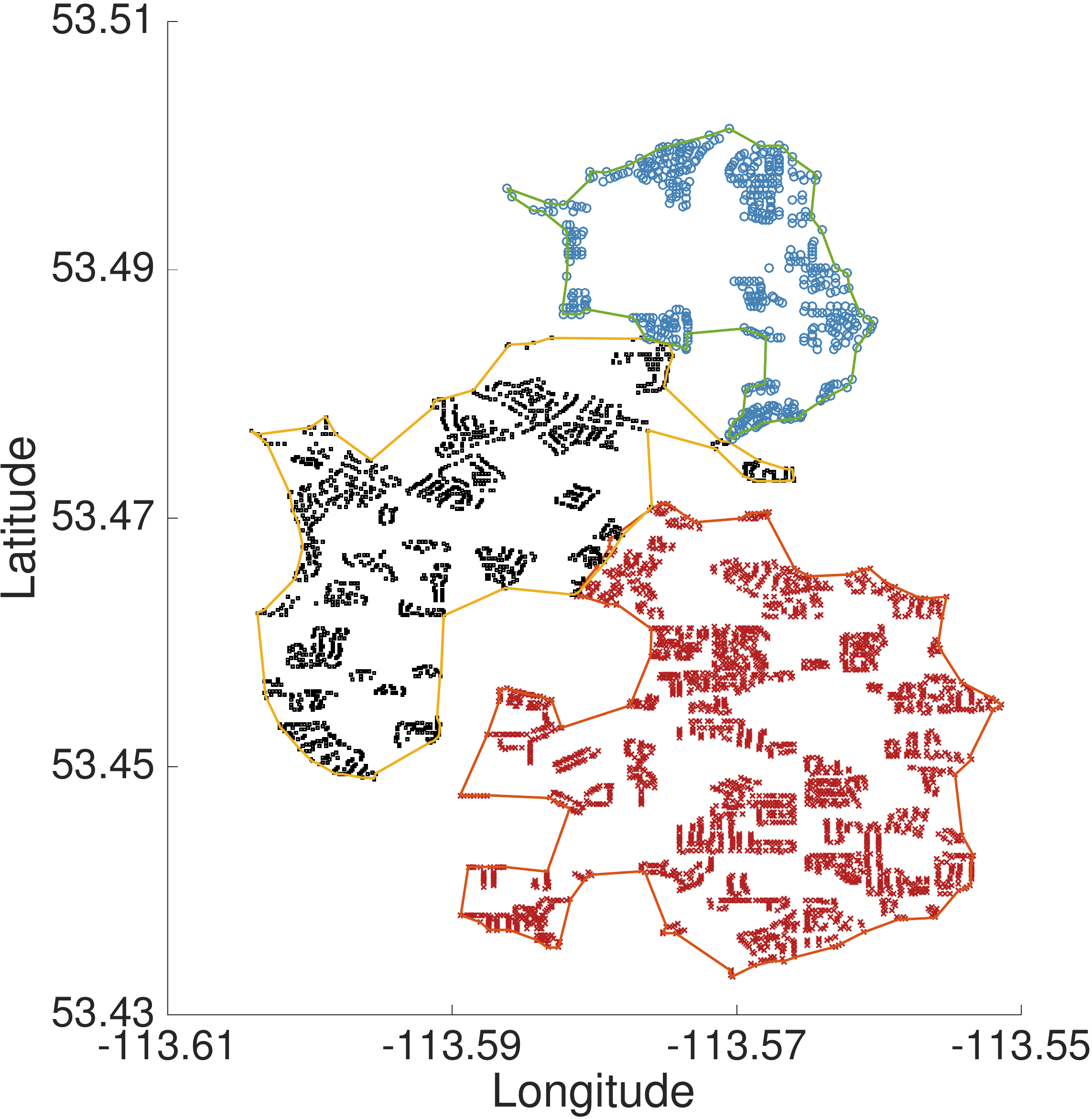}
 \caption{Three local regions found by Geo-partition, which coincide with a split by the Terwillegar Drive and the Riverbend Hill Road.}
 \label{fig:cluster3}
\vspace{-2mm}
\end{figure}

The implementation of the model \eqref{eq:hier_model} requires the spatial partition of the whole region $\Omega$ into local regions $\Omega_j$, $j=1,\ldots J$ such that $\Omega_j \cap \Omega_{j'} = \emptyset$, for $j\neq j'$,  and  $\cup_j \Omega_j = \Omega$. This partitioning is the Step 2 in Algorithm 1. In order to achieve higher prediction power, the partitioning should be based on meaningful criteria and thus the resulted partition is capable of capturing the homogeneous local regions in term of house prices. In other words, in each sub-region, the local spatial fields $f^l$ is smooth and the local attribute contributions ${\boldsymbol{\beta}}^l$'s are constant. 
In this work, we propose to use a spatial constrained clustering based on the global land desirability to partition the whole region into subregions, namely \emph{geo-partition}; see Fig.~\ref{fig:cluster3}.

In geo-partition, we propose to use a spatial constrained clustering approach based on the estimated global fields $\hat{f}^g$. In particular, 
to identify significant areas as partition patches, we combine the Penalized Spatial Distance (PSD) measure~\cite{zhang2007geo} with the Clustering by Fast Search and Find of Density Peaks (CFSFDP) clustering algorithm~\cite{rodriguez2014clustering} to partition houses based on both spatial and non-spatial attributes.
The PSD distance measure takes both spatial coordinate and non-spatial attributes into account, and is able to yield geographically separated clusters when combined with specific clustering algorithms. The calculation of PSD distance requires the non-spatial attribute's value surface. We utilize the house price as non-spatial attribute. To get a surface fitting of house prices, we apply the SSR model to the dataset with $\lambda=10^{-3}$. Then we calculate the PSD distance between each pair of houses to get a distance matrix $\mathcal{D}$.

We then apply the CFSFDP clustering algorithm on the distance matrix $\mathcal{D}$ to divide houses into geographically separated sub-regions. 
CFSFDP clustering algorithm spot clusters based on the idea that cluster centers are characterized by a higher density than their neighbours and by a relatively large distance from points with higher densities. It is able to detect variant shape clusters, and it doesn't require to specify the number of clusters. We refer to~\cite{zhang2007geo} and~\cite{rodriguez2014clustering} for more details about PSD and CFSFDP.

The resulted three clusters are shown in Fig. \ref{fig:cluster3}. The left two subregions are split by  the Riverbend Hill Road. The houses in the left top subregion were built earlier than the other two subregions and there are relatively more condos in that subregion. This means geo-partition indeed generates meaningful partition and our model will gain more prediction power by learning local structure from those subregions. 

\section{Performance Evaluation}
\label{sec:evaluation}

\begin{figure}[t]
    \centering
            \subfigure[MSE: non-log version models]{
                \includegraphics[width=1.5in]{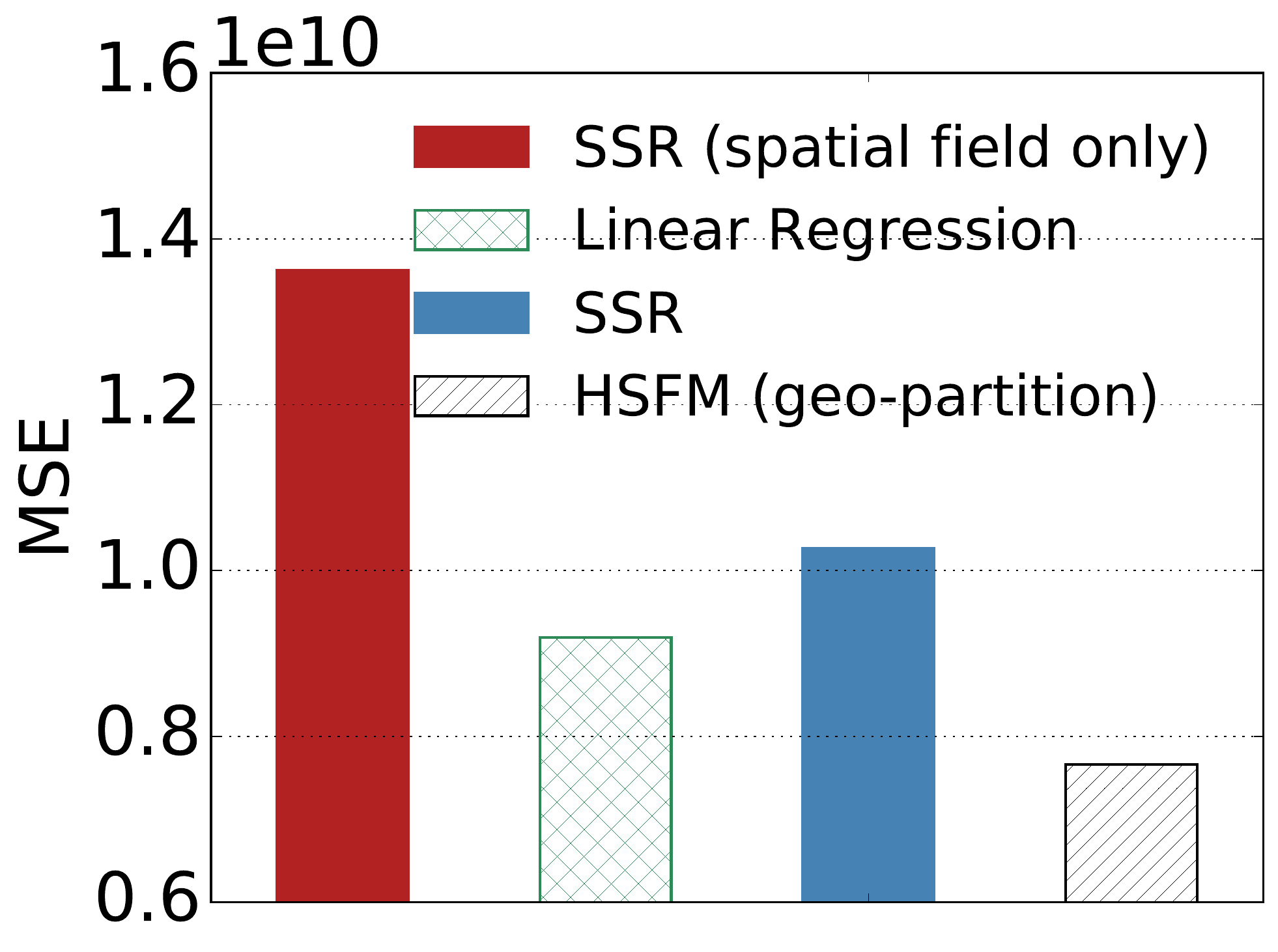}
                    \label{fig:MSE-nolog}
            }
            \hspace{-2mm}
            \subfigure[MSE: log version models]{
                \includegraphics[width=1.5in]{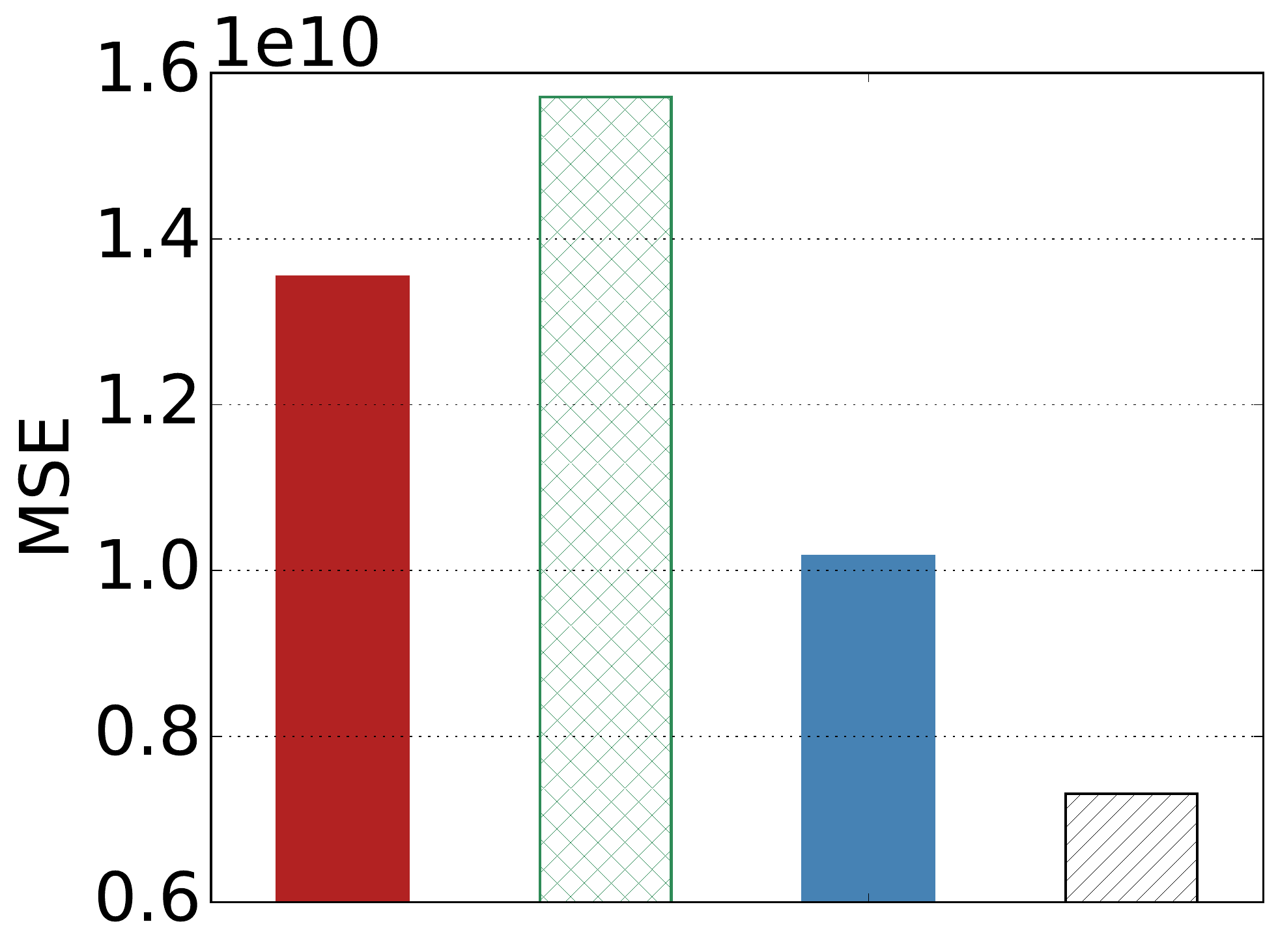}
                    \label{fig:MSE-log}
            }
            \vspace{-2mm}
    \caption{MSE of house price predictions under different methods.}
    \label{fig:compareMSEs}
\vspace{-2mm}
\end{figure}

\begin{figure}[t]
    \centering
            \subfigure[MRAE: non-log version models]{
                \includegraphics[width=1.5in]{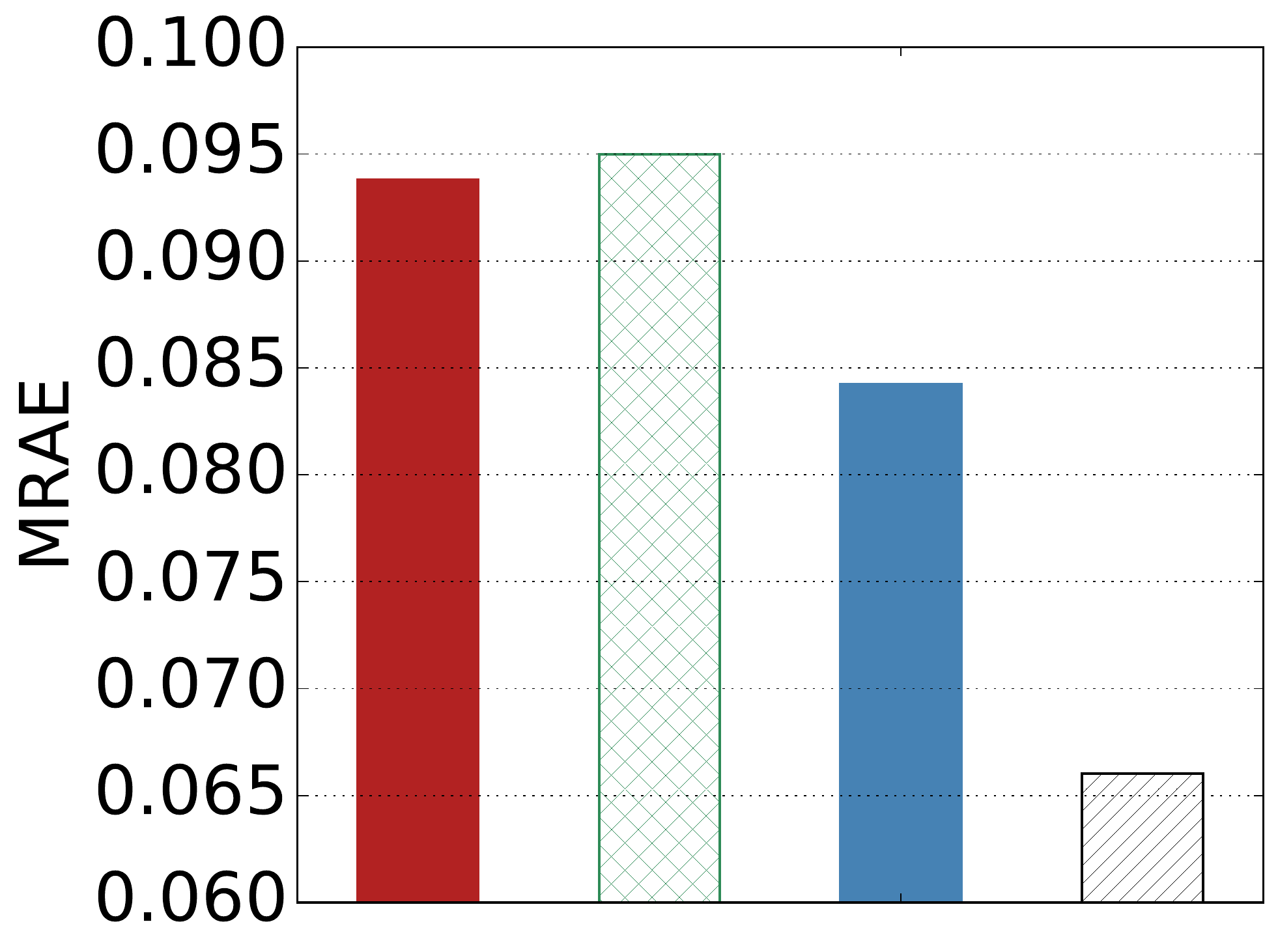}
                    \label{fig:MRAE-nolog}
            }
            \hspace{-3mm}
            \subfigure[MRAE: log version models]{
                \includegraphics[width=1.5in]{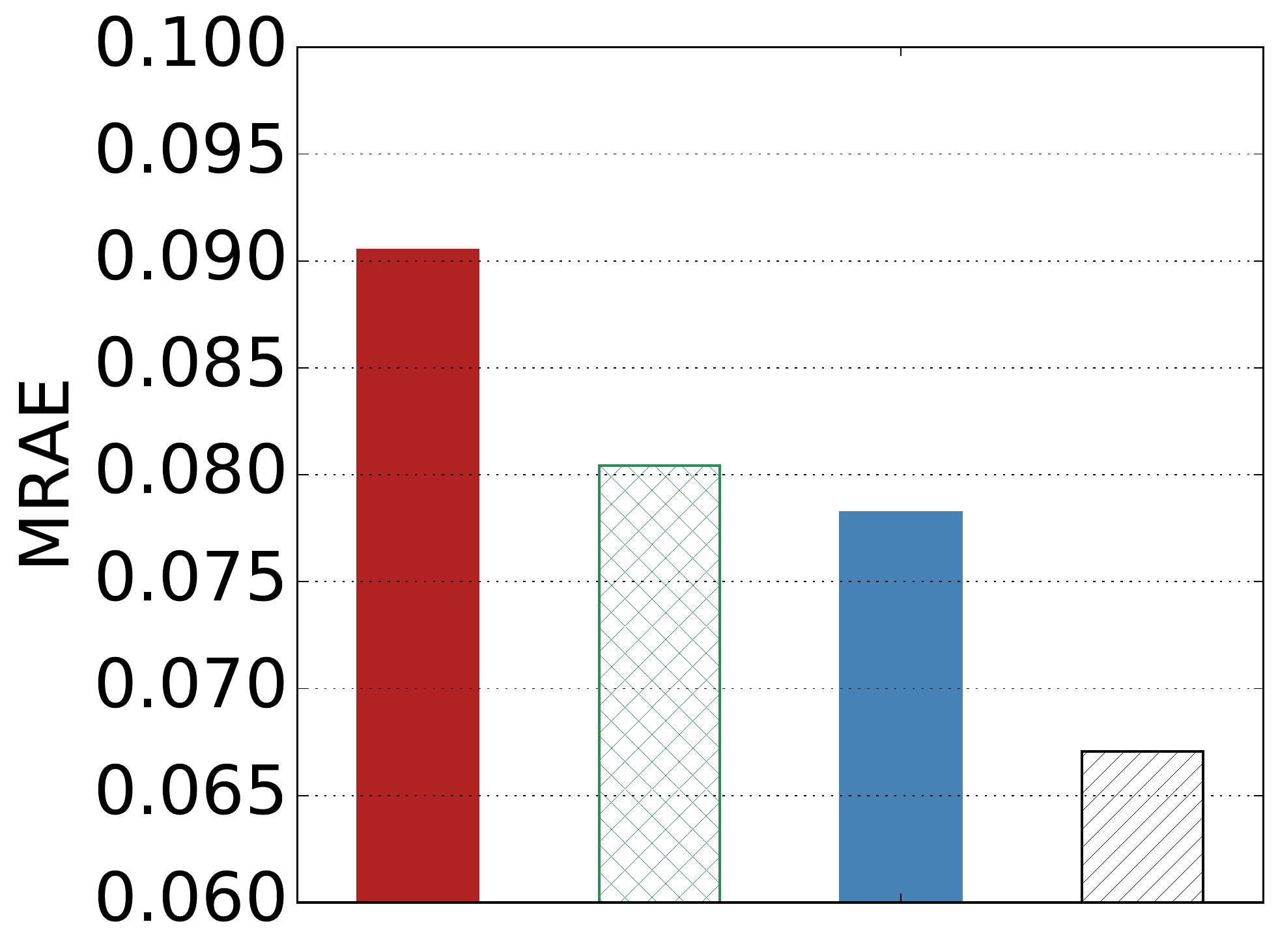}
                    \label{fig:MRAE-log}
            }
            \vspace{-3mm}
    \caption{MRAE of house price predictions under different methods.}
    \label{fig:compareMRAEs}
\vspace{-2mm}
\end{figure}


To evaluate the effectiveness our proposed approach in modeling house values in typical urban regions in the North America, we use the publicly accessible real-estate property assessment data from the City of Edmonton, Alberta, Canada, which is the 5th largest city in Canada. The dataset considered contains the assessed market values of 6130 residential houses (including single houses, duplex and townhouses) in the RiverBend area of Edmonton in 2015, together with their corresponding features including effective build year, net area, characteristics of basement, presence of a fire place, presence of air conditioning, lot size, site coverage and north/south location. The data considered is quite heterogeneous: the mean house value is CAD 655,894 and the median is CAD 598,000, while the house values range from only CAD 65,000 all the way up to CAD 2,855,500, with the heat map of house values shown in Fig.~\ref{fig:region}. The datasets and code are made publicly available \footnote{https://github.com/BangLiu/RealEstateModeling/} for future research. 

It is worth noting that the collected data points are still sparse and only represent half of the houses in the region being considered in Fig.~\ref{fig:region}. To test the generalizability of the model, we randomly sample 80\% of all houses as the training set, where we know the full information about the houses including their assessed value, features and GPS locations. 
The other 20\% of the houses serve as the test data, where only their features and GPS locations are known. The objective is to estimate the assessed values of houses in the test set. 




In particular, we evaluate and compare the following approaches on the same data:
\begin{itemize}
\item
\textbf{LR}: linear regression based on the model $z_i = \mathbf{w}_i^{\sf T}\boldsymbol{\beta} + \epsilon_i$ without the spatial field.

\item 
\textbf{SSR (spatial field only)}: spatial spline regression based on the model
$z_i = f(\mathbf p_i) + \epsilon_i$.
 \item
\textbf{SSR}: the spatial spline regression model \eqref{eq:spatial_model} \cite{Sanga13, Ramsay02} solved by finite element functional analysis.
\item
\textbf{HSFM (geo-partition)}: hierarchical spatial functional model \eqref{eq:hier_model} with spatial constrained clustering based on CFSFDP at the local scale.
\end{itemize}

For each model, we have two versions, where $z_i$ is either the original value or log-value of house $i$. We always evaluate the performance by the mean squared error (MSE) and mean relative absolute error (MRAE) of the produced estimates for the original (non-log) house value. For HSFM models, the geo-partitioning can actually be precomputed before the global and local spatial functional fitting happens.

Fig.~\ref{fig:compareMSEs} and Fig.~\ref{fig:compareMRAEs} compare the performance of baseline models and models proposed in this paper by presenting the MSEs and MRAEs given by different methods respectively. As we can see that our proposed models significantly outperforms baseline models by a great percentage, which proves the effectiveness of our models. 

Comparing HSFM with geo-partition to the first 3 baseline models, we can see that it improves a lot by introducing the secondary level model fitting on local subregions. In this way, after we fit the global trend of house price surface, the local level model fitting is able to further characterize different regions' local variance, leading to the improvement of modeling accuracy. Compare the MSEs of log version and non-log version algorithms, we can see that the log version method gives better MSE, while non-log version gives better MRAE.

\begin{table}[t]
\centering
\caption{The CDF of relative absolute errors (RAEs).}
\label{tb:cre}
\scalebox{0.85}{
\begin{tabular}{rrrrrrrr}
  \hline
 Method & $<$ 1\% & $<$ 2\% & $<$ 3\% & $<$ 5\% & $<$ 10\% & $<$ 15\%\\ 
  \hline
  \hline
LR (non-log) & 0.07 & 0.15 & 0.21 & 0.34 & 0.62 & 0.80\\ 
  SSR (non-log) & 0.08 & 0.17 & 0.24 & 0.40 & 0.69 & 0.84\\ 
  HSFM (non-log) & $\mathbf{0.12}$ & $\mathbf{0.23}$ & $\mathbf{0.36}$ & $\mathbf{0.52}$ & $\mathbf{0.79}$ & $\mathbf{0.91}$\\ 
\hline
LR (log) & 0.09 & 0.19 & 0.28  & 0.42 & 0.71 & 0.85\\ 
SSR (log) & 0.11 & 0.19 & 0.28 & 0.45 & 0.73 & 0.86\\ 
  HSFM (log) & $\mathbf{0.12}$ & $\mathbf{0.24}$ & $\mathbf{0.36}$ & $\mathbf{0.53}$ & $\mathbf{0.79}$ & $\mathbf{0.90}$\\ 
   \hline
\end{tabular}
}
\vspace{-4mm}
\end{table}

As a summary, we compare the CDF of relative absolute errors (RAEs) under different methods in Table~\ref{tb:cre}. We can see that HSFM (log) yields the best performance with $53\%$ of estimates having an error less than $5\%$, proving the superior performance of the new model.

\section{Concluding Remarks}
\label{sec:conclude}

We propose the Hierarchical Spatial Functional Model (HSFM) for house price modeling based on the recent advancement in spatial functional data analysis. 
It models house prices as a combination of a global spatial field, representing the global-scale land desirability surface, and multiple local spatial fields, each characterizing a local-scale land desirability surface, as well as a linear regression from house features. We propose effective methods to discover the hidden homogeneous sub-communities that may exist in a heterogeneous region of interest based on spatial constrained clustering. We then solve the proposed HSFM by applying a recently developed finite-element analysis technique called spatial spline regression on both the global and local levels in sequence. We demonstrate the effectiveness of our model for house price estimation based on partially known house prices based on more than 6000 houses in Edmonton, Alberta, Canada.

\bibliographystyle{IEEEtran}
\bibliography{main}

\end{document}